# Propagation of Wave Packets in Dispersive Media

J Ernest Breeding, Jr.*

Naval Research Laboratory
Stennis Space Center, MS 39529 USA

*Now unaffiliated (retired)
115 Blackbeard Dr
Slidell, LA 70461 USA

sunny38@caa.columbia.edu

## Abstract

The propagation and refraction of waves in dispersive media are considered.  A primary objective is to determine whether waves refract as monochromatic waves by Snell's law with phase velocity or as wave packets.  The refraction of wave packets requires two refraction laws.  The wave packets refract according to Snell's law with the geometric group velocity while the wavelets within the wave packets refract according to Snell's law with phase velocity.  The tests were performed using directional power spectra from gravity water waves originating in the Caribbean Sea, Gulf of Mexico, Indian Ocean, and the Southern Ocean.  In all cases the waves were measured with arrays of pressure sensors.  It was found that the refraction of monochromatic waves led to backtracks from a measurement site that spread out for the different wave periods and did not go back to a common source.  When the refraction was large the results were especially bad.  When the data were used to test the refraction of wave packets very good results were obtained.  For every storm considered the backtracks of the different wave periods went back together to the wave generation area. These results offer strong support for refracting waves in dispersive media as wave packets, not as monochromatic waves. This result for wave groups was also determined in a controlled wave tank experiment of gravity water waves where a critical angle was tested. The properties of propagating wave packets are also considered and illustrated. Reflection points and critical angles are found and explained. There can be major differences between the propagation of wave packets and monochromatic waves.

## 1. Introduction

In section 2 the history and derivation of group velocity in the late 1800s by Stokes (1905) and Rayleigh (1877) are discussed. In dispersive media the wave pattern for an instant of time differs from the pattern for a constant position. The difference is related to the ratio of the group velocity to phase velocity (Breeding, 1967). Snell's law for monochromatic waves dates from the mid 1600s. An expression for the ray curvature for monochromatic waves was derived by Munk and Arthur (1952).

In section 3 Peak (2014) clearly shows that the accepted method for the bathymetric refraction of wave packets is to use Snell's law with phase velocity. Further, Munk et al (1963) determined backtracks to the Southern Hemisphere from an array of three pressure sensors off the coast of La Jolla, California. But their results based on bathymetric refraction using Snell's law with phase velocity missed the storms and were in error. Munk et al (2013) published a correction to the earlier paper based on the work of Gallet & Young (2014) that attributed the errors to refraction by the vorticity of surface currents near the equator.

Section 4 has a description of the instrumentation for two arrays with six pressure sensors in the Gulf of Mexico near Panama City, Florida. Directional power spectra were determined from the data.  In section 5 backtracks were computed for data from Hurricane Betsy of September 1965.  Bathymetric refraction was based on Snell's law with phase velocity.  The results were really bad missing the storm completely (Breeding, 1972).

In section 6 since travel times are determined using group velocity it is logical to refract wave packets in dispersive media using group velocity in Snell's law. However, it is the wavelet direction that is determined at the measured site.  The wave packet direction can be determined by the refraction-source method (Breeding, 1978).  It is assumed that the direction of a wave packet and its wavelets are equal where the waves are generated.  Then, using both Snell's laws for the wave packet and Snell's law for the wavelets the equations can be interrelated to solve for the wave packet direction at the measurement site.  The water depth at the wave source is needed to compute the velocities, and that can be found by trial and error.  Backtracks were determined in section 7 where the path was computed using the ray curvature expression for group velocity while at each point of the path the wavelet directions were determined by Snell's law with phase velocity.  Two refraction laws are required.  The results were very good.  (Breeding, 1972).  Black (1979) had directional wave data from Hurricane Eloise that passed through the Gulf of Mexico in September 1975.  For the refraction of wave packets he followed the method used by Breeding (1972).  Black got good results for refraction based on group velocity but not for monochromatic wave refraction.

In section 8 it is stated that the velocity for wave packets propagating in dispersive media is the geometric group velocity.  It is defined as the product of the conventional group velocity and the cosine of the angular difference between the wave packet and wavelet directions.  Snell's law with the geometric group velocity and the corresponding ray curvature expression are given.  The computer program used to determine wave packet trajectories is also described.

In Section 9 the direction-wave number method is presented for determining the wave packet directions from the measured wavelet directions.  The method is based on the variation of the wavelet directions in radians versus the wave numbers (Breeding, 1980).  Also, wave packet backtracks with the geometric group velocity were computed using the two refraction laws for waves measured with the Panama City array for Hurricane Fifi.  This storm passed through the Caribbean Sea in September 1974. The wave packet backtracks were good as they went back through the Yucatan Channel to the storm.  Monochromatic backtracks were not good as they missed the Yucatan Channel and could not reach the generation area (Breeding, 1978).  As a further test of the refraction laws Tang (1994) produced backtracks for the Munk et al (1963) data for waves from distant storms.  The wave packet backtracks using the two refraction laws went back to the storms for all wave periods considered.  But the backtracks for monochromatic waves were bad as they spread out in many directions and did not go back together for any of the storms.

Section 10 is about the attributes of propagating wave packets.  The interesting properties of the wave packet ray curvature expression are illustrated.  For wave packets propagating from deep water there is a critical angle of 74.8 degrees (Breeding, 1980).  The wave packets are turned parallel to the water depth contours.  It is shown that the conventional group velocity can be a good approximation to the geometric group velocity.  A reflection point will occur for wave packets propagating towards deep water if the wavelet direction is turned parallel to the water depth contours (Breeding, 1980).  The relationship between Fermat's principle, the Euler-Lagrange equation, the ray curvature equation, and Hamilton's equation are explained (Breeding, 1986).  When the water depth contours and the shoreline are sinuous, wave packets propagating from deep water have the largest wave heights in the bays, where land is missing, and the smallest wave heights at the headlands, just the opposite of what happens for monochromatic wave refraction (Breeding, 1981).

Section 11 presents the results of a controlled wave tank experiment (Linzell, 1987).  Wave groups were generated by a paddle wheel in deep water at an angle near the critical angle of 74.8 degrees.  The wave heights decreased rapidly going from deep water to shallow water in the tank.  This is not consistent with monochromatic wave refraction.  This experiment provides strong support for the refraction of wave packets according to the two refraction laws.  Section 12 has the conclusions of the article and section 13 the references.

Gravity water waves are used to illustrate the propagation of waves and their properties for different bathymetry.  However, many of the concepts discussed apply to waves in general.  For example, seismic surface waves are dispersive.  The refraction results presented here will apply to them as well as other kinds of dispersive waves.

## 2. Origin of Group Velocity, Definitions, and Wave Patterns

## 2.1. Group Velocity and Definitions

During the latter part of the nineteenth century there was interest in predicting when water waves from distant storms would arrive at the England coast.  Researchers were using the phase velocity of gravity water waves to predict the time of arrival.  But it was taking longer for the waves to arrive than they predicted.  An answer was sought for this discrepancy.

It was known that waves from distant storms arrived in groups separated by regions of calm.  These kinds of waves are known as swell.  For dispersive waves it was known that a wave group moves at a different velocity than the phase velocity.  William Froude made this observation in his experimental studies of gravity water waves.  He wrote of his findings to Sir George Gabriel Stokes on January 17, 1873 (Stokes, 1907).  Stokes found a mathematical relationship between the group velocity and phase velocity based on wave interference. It was published as a Smith's prize examination paper dated February 2, 1876 (Stokes, 1905}. Froude also informed Lord Rayleigh of his observations of wave groups. Working independently of Stokes, Rayleigh (1877) arrived at the same explanation and mathematical relationship for the group velocity.

Stokes and Rayleigh obtained their result by summing two component sine waves of equal amplitude, nearly equal frequencies, and moving in the same direction. Breeding (1978) presents this derivation and also considers a more realistic wave packet with a narrow but continuous range of frequencies, variable amplitudes, and with the waves moving in a narrow range of directions.

Purser and Synge (1962) and Synge (1962, 1963) recommended that water waves should have a name like other waves such as photons and gravitons. They called water waves hydrons. We will use that term for properties that apply only to gravity water waves.

The phase speed of a hydron is given by (Lamb, 1932) as

$$v = \left[\frac{g}{k}\tanh(kh)\right]^{1/2} \tag{1}$$

where g is the acceleration due to gravity, k is the wave number, and h is the water depth. When $h/\lambda > ½$, where $\lambda$ is the wavelength, the water depth is defined as deep water. Then equation 1 reduces to

$$v_d = \frac{gT}{2\pi} \tag{2}$$

where T is the wave period. For all practical purposes when h/λ < 1/20 the water depth is defined as shallow water. Then equation 1 reduces to

$$v_s = \sqrt{gh} \tag{3}$$

The only variable in this equation is the water depth.

The group speed U for a hydron is defined by (Lamb, 1932) as

$$U = \left[1 + \frac{2kh}{sinh(2kh)}\right]\frac{v}{2} \tag{4}$$

In deep water U reduces to

$$U = \frac{v}{2} \tag{5}$$

This is the group speed that Stokes and Rayleigh derived to determine the travel times of waves. In shallow water we have

$$U = v \tag{6}$$

In deep water and intermediate water depths v and U are dispersive and in shallow water they are not dispersive. Group velocity and phase velocity differ in dispersive media.

## 2.2. Wave Patterns in Dispersive Media

Following Stokes and Rayleigh, Breeding (1967) compared the interference patterns for non-dispersive and dispersive media by summing two sine waves of equal amplitude, of nearly equal frequency, and moving together in the same direction. Consider the wave patterns between alternate group maxima. If the waves are non-dispersive the wave pattern with time for a constant position is the same as the wave pattern that varies with distance for a constant time. For a dispersive media these wave patterns are not the same. For the example of hydrons the non-dispersive case is comparable to shallow water where the conventional group velocity and phase velocity are the same. For the dispersive case of waves in deep water the pattern considering the variation with distance for an instance of time has only one half as many oscillations as the pattern that varies with time for a constant position. This ratio of one half corresponds to the ratio of group velocity to phase velocity for waves in deep water. See Breeding (1967) for more details and the derivation of a periodicity ratio for the wave patterns that applies to any kind of dispersive waves.

## 2.3. Snell's Law and Ray Curvature

In 1621 Willebrord Snell discovered experimentally the refraction law that bears his name. He was studying the propagation of light from one media to another. However, Snell did not publish his result. It was first published in 1637 by Rene Descartes (Born & Wolf, 1965). Snell's law is stated

$$\frac{sin\gamma}{v} = Constant \tag{7}$$

where $\gamma$ is the angle of a monochromatic wave. For ray tracing in media where v varies it is easier to use an expression for the ray curvature instead of Snell's law. An expression was derived by Munk & Arthur (1952) and Arthur, et al (1952)

$$\kappa = \frac{1}{v}\left(sin\gamma\frac{\partial v}{\partial x} - cos\gamma\frac{\partial v}{\partial y}\right) \tag{8}$$

where $\kappa$ is the ray curvature. Equation (8) can be integrated to obtain equation (7).

# 3. Wave Data Measured in the Pacific Ocean near La Jolla, California

## 3.1. Bathymetric Refraction - The Established Method for Wave Packets

In an article by Peak (2004) the interest is the accuracy of refraction when there is complex bathymetry. The study area was two submarine canyons off the coast of La Jolla, California where long period hydrons undergo strong refraction. Their predictions were compared with three months of data collected in the Fall of 2003 with 7 directional wave rider buoys, 17 bottom pressure recorders, and 12 pressure-velocity sensors. They found a difference between predictions and measurements of about 20%.

It is clearly shown that the bathymetric refraction was determined according to Snell's law with phase velocity. The wave packets follow the paths of monochromatic waves. This is and has been the accepted way to determine the directions and paths of refracted wave packets.

## 3.2. Backtracks to Distant Storms

In a milestone paper Munk, et al (1963) describe the measurement and directional wave analysis of water waves from distant storms. The waves were measured using a triangular array of pressure transducers located at a water depth of 100 m 3.2 km offshore from San Clemente Island in the Pacific Ocean west of San Diego, California.

The pressure sensors were placed at the corners of what was approximately an equilateral triangle. The separation distances between the sensors were 272, 282, and 296 m. Cables carried the signals to a telemetering tower on San Clemente Island where the signals were transmitted to a receiving tower at La Jolla. From there telephone lines carry the signals to the laboratory where the data were digitally recorded.

Data were collected at intervals of every 4 seconds, or a frequency of ¼ Hz. The Nyquist frequency is 1/8 Hz or 125 cycles per kilosec where folding of energy occurs. Due to the depth at which the sensors were placed no aliasing was expected due to the water column acting as a low pass filter.

The covariance was computed from the time series records. The Fourier transform of the covariance gives the directional power spectra. From these data they were able to identify storms by plotting standard energy levels on a plot of frequency versus time. The contours so obtained show pronounced slanting ridges. The slanting ridges are associated with dispersive arrivals from individual storms. The slope of a straight line drawn through a slanting ridge determines the distance to the storm. The intercept on the axis (where frequency is zero) is the time of origin of the storm.

With the triangular array Munk et al (1963) collected wave data from June through November in 1959. They selected 30 storms to study. In order to infer the locations of the storms they had to account for refraction due to bathymetry in the vicinity of the pressure array. But they did not do an adequate job. As a result, they focused mostly on the higher frequency waves in the spectra since they were the least affected by refraction. To refract the wave packets they used Snell's law with phase velocity for monochromatic waves.

Most of the storm waves came from great distances and were from the New Zealand-Australian-Antarctic region or the Ross Sea. The inferred storm locations were compared with the actual storms on weather maps. They measured antipodal swell from the Indian Ocean nearly half way around the world. Three of the inferred storm locations were found on the Antarctic Continent. The inferred locations were in error by about 15 degrees compared with the actual storms in the ocean on weather maps.

Munk et al (1963) were disappointed in the results since they did not correctly go back to the wave sources. In the southern hemisphere they should go back to the right of a storm center due to the clockwise wind motion. That did not happen. They stated: "There is a curious indication that the wave-inferred directions are to the left of the location obtained from weather maps." They also stated refraction should be given more careful consideration. This will be discussed further in Section 9.3.

## 3.3. Backtracks to Distant Storms – Correction

Munk et al (2013) published a correction to their article of Munk et al (1963). The correction was based on a proposal by Gallet & Young (2014) that refraction by currents was responsible for the errors in the inferred directions to some of the storms. (Note that an earlier version of the 2014 paper had been seen by Munk et al (2013).)

Gallet & Young (2014) used pitch and roll data collected at a location not far from where the triangular array of Munk et al (1963) was located. Their objective was to use modern data from storms from the Southern Hemisphere to compare with the findings of Munk et al (1963). To test the refraction by currents proposal they used data from the OSCAR program to compute the Vorticity of eddy currents near the equator. Waves from the storms would pass through the eddy currents before reaching the measurement site.

Gallet & Young (2014) were of the opinion that most of the Munk et al (1963) backtracks would follow great-circle paths. But for the few inferred sources that were incorrect the trajectories had been deflected from great-circle paths by the refraction due to the vorticity of surface currents. This includes the storms that had inferred locations on the Antarctic Continent. These locations were said to be in error by as much as 10 degrees due to current refraction.

Gallet & Young (2014) pointed out that the larger wave periods that have group velocities much larger than the currents, which are usually less than 1 m/sec, would not be deflected from great-circle paths. It is the waves with the higher frequencies that are deflected away from great-circle paths due to current refraction.

Following Gallet & Young (2014) Munk et al (2013) show that by accounting for a deflection of 10 degrees by current refraction they can move their inferred sources from the Antarctic Continent to the ocean where the storms actually occurred. They point out that waves with a low frequency of 0.03 Hz have a group velocity of 26 m/sec which is much higher than the current velocity, and therefore current refraction can be ignored as stated above. These low frequency waves are the first waves to arrive at the measurement site. Waves with a frequency of 0.07 Hz have a group velocity of 11.4 m/sec, and they will be deflected or scattered by currents. These higher frequency waves take longer to reach the measurement site. Some of the scattered waves due to current refraction, and which do not follow great-circle paths, do not reach the measurement site because they are blocked by the Cortez bank. So, scattering due to current refraction can have a major impact on the directions of the measured waves. This will be discussed further in Section 9.3.

## 4. Wave Data Measured near Panama City, Florida

## 4.1. Instrumentation - Pressure Arrays

The instrumentation for data collection is described by Breeding (1972) and summarized here. In the Gulf of Mexico near Panama City, Florida there once were two offshore platforms. They were used for research by the U.S. Navy. Stage I was located in 31.7 m of water about 17.7 km from shore. Stage II was located in 19.2 m of water about 3.2 km from shore. On the sea floor near each stage there was an augmented pentagonal array of six pressure sensors for measuring gravity water waves from storms.

The arrays at each stage were identical.  A sensor was placed at each corner of a pentagon with sides of 35.8 m.  The sixth sensor was placed at the center of the array 30.5 m from the other sensors.  The arrays were designed by Bennett, et al (1964)

The configuration of the array is obviously not the best way to place six sensors to measure wave directions from data.  But the sensors cannot be depended upon to work all of the time.  With the configuration chosen if one or several sensors were not working the directional resolving power of the arrays did not degrade much.

Measured output data from each sensor was fed by a cable to electronic equipment on the stages.  After signal processing the data from Stage I was sent by radio telemetry to a receiver at a beach tower on shore.  The stage II data was fed by a submarine cable to the beach tower.  At the beach tower there was more processing of the data and it was recorded on separate tracks of a magnetic tape recorder.  After computer processing of the data a library tape was created such that wave data for specific storms and time periods could be retrieved in order to do directional power spectral analysis of the data.

## 4.2. Directional Power Spectra

The method for computing directional power spectra from the data is presented by Breeding (1972) and summarized here. Bennett et al (1964) and Bennett (1968) computed the directional power spectra following Munk. et al (1963).  The data was sampled at intervals of one second.  That means the Nyquist, or folding frequency, was 0.5 Hz.  Since the pressure sensors are at depth there was very little if any energy above 0.5 Hz measured.

For each pressure sensor a wave record of about 30 minutes or 1800 data points were recorded for spectral analysis.  After computing the covariance and taking the Fourier transform of it the directional power spectra is obtained.  With regard to accuracy of the calculations each spectral estimate has a resolution of $\Delta f = 1/120$ Hz.  There are 60 degrees of freedom.

For an example of the computed directional power spectra wave data from Hurricane Betsy will be considered (Breeding, 1972). Hurricane Betsy passed over the southern tip of Florida and entered the Gulf of Mexico a little prior to noon on September 8, 1965.  After making an arc it entered land at the Mississippi Delta just before noon on September 9, 1965.  The waves were generated in intermediate water depths.  In Figure 1 is the directional power spectra for September 9 for the time 0759-0830 Central Standard Time (CST).  In the figure the frequencies from about 0.00 to about 0.15 Hz are of interest. Due to the consistency of the data in this frequency range these waves are obviously from a single storm.

## 5. Backtracks according to Snell's Law with Phase Velocity

To trace the rays (Breeding, 1972) for monochromatic rays moving with phase velocity the computer program of Wilson (1966) was used.  The water depth grid for the area off of Panama City was created by reading water depths off of U.S. Coast and Geodetic Survey Charts.  The square grid of water depth values had a spacing between grid points of 4.537 km. Each point of a ray is determined using the ray curvature equation (8).  At each ray point a plane is fitted to the four closest water depth grid points by the method of least squares. The slope of the plane determines the gradients of the water depths which are used to determine the velocity gradients.  Also, the water depth value is estimated.

Figure 2 shows the backtracks determined from the measure site at Stage I for the directional power spectra presented in Figure 1. The rays completely miss the path of the storm.  This result was unexpected and very surprising and is very important because of the ramifications that follow from it. The backtracks of this refraction diagram have been checked and checked and are thought to be correct. Then what is the correct way to refract waves?

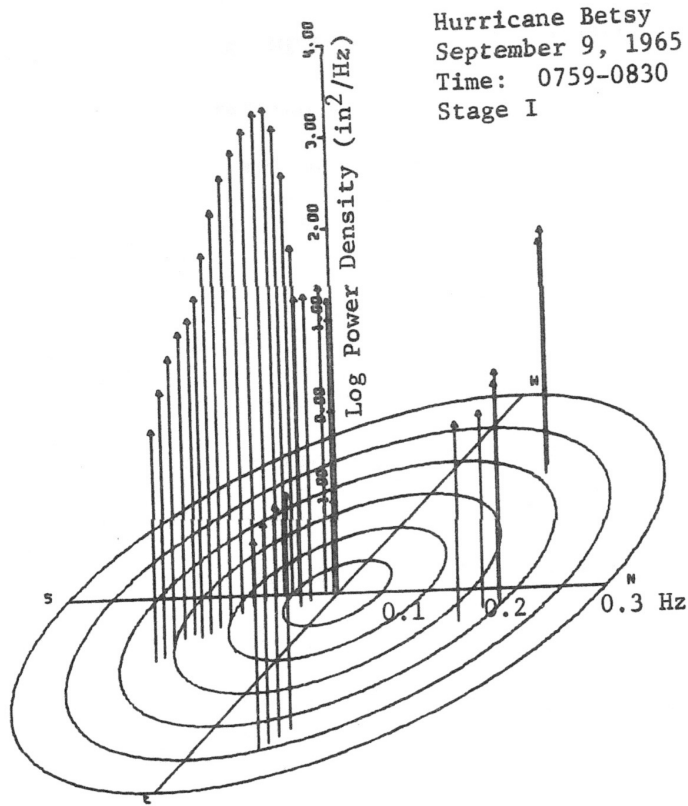

**Figure 1:** An example of directional power spectra. The bottom of an arrow gives both the frequency and bearing of the waves. Note that the frequency increases in the form of concentric circles moving outward from the center. The direction north (N) represents magnetic north. To obtain true north add 3 degrees. The tip of the arrow indicates the log power density of the waves at the depth of measurement. The time is Central Standard Time.

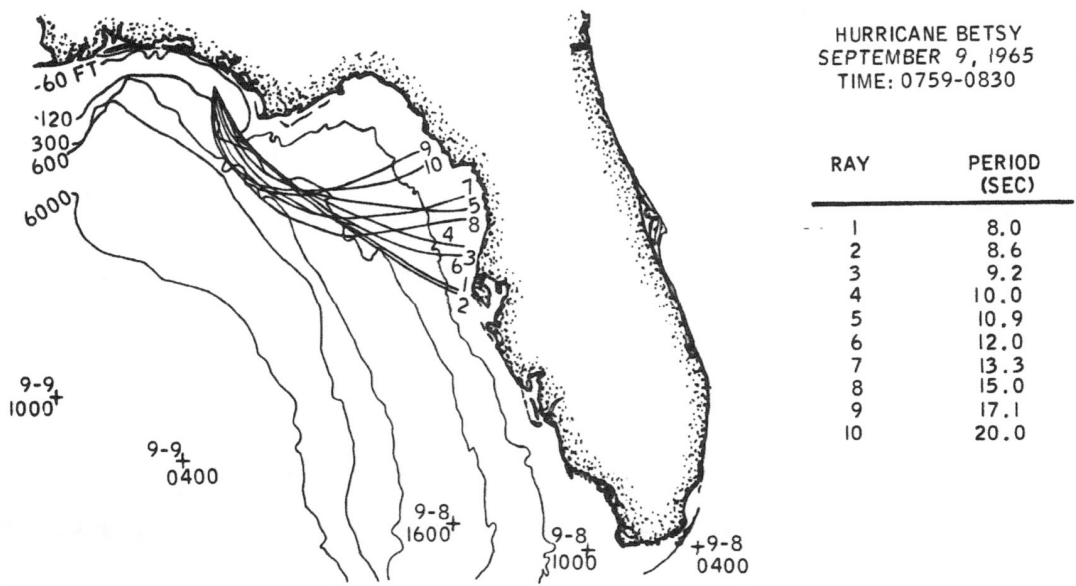

**Figure 2:** Monochromatic ray trajectories from Stage I. The wave periods of the rays are shown. The path of Hurricane Betsy is shown on the bottom of the figure. At each location the month-day and 24-hour time are given. The time is Central Standard Time. The water depths are in ft. 1 m equals 3.28 ft.

## 6. Snell's law with group velocity - two refraction laws?

## 6.1. Group Velocity versus Phase Velocity

It seems that there are three important things to consider in refracting water waves due to bathymetry. Group velocity is used to predict the travel time of waves (section 2.2). Secondly, when you consider gravity water waves the groups contain wavelets that move through them at the phase velocity of the waves. The wavelets start at one side of the group and race though dying out at the other side.  The wavelets do not persist.  It is the wave groups that persist.  The third consideration is Fermat's principle (Breeding, 1986).  You want to find the path of least time for a ray to move from point A to point B.  Then would it not make sense to use group velocity in Fermat's principle?

It was decided that the refraction of wave packets should be considered.  But to do that it was necessary to develop a program for refracting wave packets, a more complex program than for monochromatic waves.  In the program the wave packets refract according to Snell's law with group velocity while at each point of the trajectory the wavelets refract according to Snell's law with phase velocity.  Two refraction laws are required.

To compute the rays for wave packets moving with the group velocity **U** a modified version of the Wilson (1966) program was used.  (Strictly speaking the geometric group velocity **G**, which is introduced in section 8, should be used.  But the use of **U** provides a good approximation to the more complex calculations using **G**.)

## 6.2. Determining the Hydron Bearings at the Measurement Site

From measurements of gravity water waves it is the wavelet bearing that is determined.  The hydron bearing has to be determined as described below.

### 6.2.1. Refraction-Source Method

In the refraction-source method Breeding (1972, 1978, 1996) assumes that where the waves are generated the hydrons and wavelets have the same direction.  Then Snell's law for hydrons and Snell's law for the wavelets can be combined and solved for the hydron bearing.  The result is

$$sin\theta_m = \frac{v_s}{v_m}\frac{U_m}{U_s}sin\gamma_m \qquad (9)$$

where $\theta$ is the hydron bearing, $\gamma$ is the wavelet bearing, the subscript m refers to the measurement site, and the subscript s refers to the water depth of wave generation, the source of the waves.

To evaluate equation (9) it is necessary to know the water depth of wave generation so that we can determine the velocities there. The water depth of wave generation is found by trial and error. Backtracks with different assumed hydron bearings at the measurement site are determined for a given wave period. At the generation site the hydron and wavelet bearings will be equal. In practice you look for the water depth where the difference between the hydron and wavelet bearings has a minimum value close to zero. Once you have tried several different trial hydron bearings you have a pretty good idea as to what to try next. Since the wave generation water depth can be different for different wave periods the method must be repeated for all wave periods of interest. The results must be consistent for the different wave periods. If the track of the storm is known this information will greatly simplify the trial and error process. Also, the travel time for the rays should be consistent with the history of the storm.

The refraction-source method is described and illustrated by Breeding (1996) for storm generation depths in deep water and intermediate water depths. An example is also presented showing how the method is used to track the movement of a storm in intermediate water depths.

## 7. Refraction of Hydrons using the Conventional Group Velocity U

### 7.1. Backtracks from Hurricane Betsy

For a number of reasons the Hurricane Betsy data offers a good test of refraction theories. The paths for which refraction takes place are long extending along the entire western coast of Florida. The angles of refraction are large. Differences between the monochromatic and wave packet refraction theories are emphasized.

The generation depths of the storm were determined by Breeding (1972) for several data sets at different times. Comparing the results, it showed that the storm was moving west from the Florida coast and into deeper water. These results would have been determined even if the track of the storm had not been known.

The backtracks from Stage I obtained for the refraction of hydrons are shown in figure 3. It is seen that the wave packet trajectories go right back to the storm. The computed generation water depths are shown on the figure and they agree with the path of the storm. The longer period waves travel faster and show you where the storm was most recently. Further, the tick marks at 5-hour intervals show that the trajectories agree with the movement of Hurricane Betsy. The conclusion: In dispersive media it is necessary to refract wave packets according to Snell's law with the velocity of the group while at each point of the trajectory the wavelet direction is determined by Snell's law with phase velocity.

## 7.2. Backtracks for Hurricane Eloise

Black (1979) analyzed direction energy spectra from measurements of waves generated by Hurricane Eloise when it was in the Gulf of Mexico September 21-23, 1975. He also considered the refraction of the waves comparing the results for both monochromatic waves and hydrons. For the refraction of hydrons he followed the method used by Breeding (1972). Black got good results for hydron refraction with the two laws but not for monochromatic refraction. He stated "This strongly supports Breeding's contention that wave energy refracts according to Snell's law and group velocity."

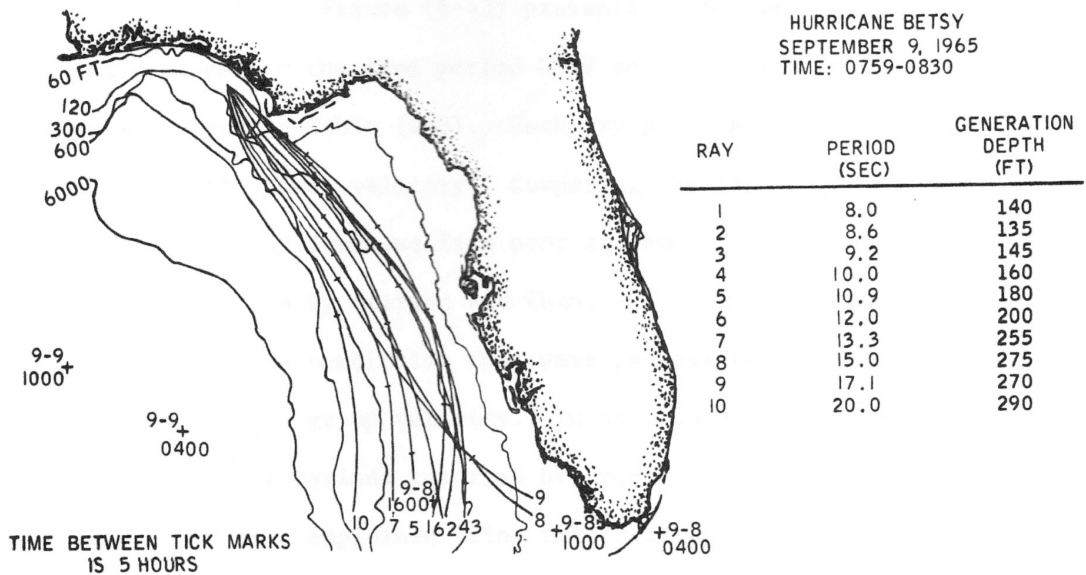

**Figure 3:** Hydron trajectories from Stage I. The wave periods of the rays are shown. The generation water depths determined by trial and error are also shown. The path of Hurricane Betsy is shown on the bottom of the figure. At each location the month-day and 24-hour time are given. The time is Central Standard Time. The water depths are in ft. 1 m equals 3.28 ft.

## 8. Geometric Group Velocity

## 8.1. Definitions

Breeding (1978) considers both wave groups formed by two sine waves moving in different directions and a continuous spectrum where the frequency spread is narrow and the component waves move in different directions for dispersive waves. It is found that the speed G of a wave packet is given by

$$G = U\cos\phi \tag{10}$$

where

$$\phi = \theta - \gamma \tag{11}$$

It can be seen that G = U only when the wave packets and wavelets have the same direction. The definition of **G** has an angular dependence, as a result Breeding (1973) proposed that **G** be called the geometric group velocity. This way it will not be confused with **U** which is well known as the group velocity.

However, as will be shown in section 10.2 the use of **U** in place of **G** is often a very good approximation to the correct answer, as it was in the case of determining the backtracks for Hurricane Betsy in section 7.1. It also leads to many simplifications in the equations for calculating expressions like the ray curvature.

Considering both a two-component wave group and the more realistic wave packet defined by a narrow band of frequencies Breeding (1978) derives Snell's laws for both the wave packet and the wavelets within the wave packet. Since the spectrum for a wave packet is a narrow spectrum Snell's law for the wavelets in all practical purposes is the same as for monochromatic waves. For the wave packet

$$\frac{\sin\theta}{G} = Constant \tag{12}$$

The ray curvature expression $\kappa_G$ for a wave packet is given by Breeding (1980)

$$\kappa_G = \frac{1}{G}(\sin\theta \frac{\partial G}{\partial x} - \cos\theta \frac{\partial G}{\partial y}) \tag{13}$$

## 8.2. Computer Program for Wave Packet Refraction

An advanced computer program for refracting hydrons has been developed. Parallel water depth contours are not assumed. From the water depth grid 12 values of water depths about a ray point are selected. A quadratic surface is fit to the chosen water depths. The water depth and its first order and second order partial derivatives are determined at a ray point in a fixed xy-coordinate system by interpolation on the quadratic surface. The first order partial derivatives of phase velocity are determined in the fixed xy-system.

At each ray point calculations are also made in a variable x'y'-coordinate system. The positive x'-axis is taken in the direction of the gradient of the water depths. This leads to simplified expressions. The first order partial derivatives with respect to y' vanish and the second order derivatives with respect to y' are simplified. The angle $\alpha$ by which the x'-axis is rotated with respect to the x-axis is computed. The value of $\phi$ and the geometric group velocity are computed.

The wavelet direction at each ray point is computed using Snell's law with phase velocity in the variable x'y'-coordinate system and then is converted to its value in the xy-coordinate system. The ray curvature, equation (13), is used to compute the path of a ray point by point. It is computed in the x'y'-coordinate system where the partial y-derivative is zero. The computed wave packet direction is then converted to its value in the fixed coordinate system.

Breeding (1986) derives equations for determining the modification to the wave height point by point along a ray. The shoaling coefficient accounts for the change in the geometric group velocity, the refraction coefficient accounts for the change in the separation distance between adjacent rays, and the friction coefficient accounts for the loss in energy. These coefficients are computed in the primed coordinate system. Since the wave height calculations involve second order partial derivatives the x'y'-coordinate system leads to great simplifications in the calculations. Other ray parameters computed in the primed system are determined. This program has been used extensively for computing the refraction of propagating hydrons.

## 9. Refraction of Hydrons using Geometric Group Velocity

## 9.1. Direction-Wave Number Method

In section 6.2.1 the refraction-source method for determining the bearings of wave packets at the measurement site was described. An alternative method is the direction-wave number method. If the data is sufficiently good this is an excellent way to determine the wave packet bearings at the measurement site. It is based on the formula derived by Breeding (1978)

$$tan\phi = k\frac{d\gamma}{dk} \tag{14}$$

A good example of the method is seen in Breeding (1980) presented in Figure 4. The wavelet bearings in radians are plotted against the wave numbers in 1/meters for datasets for two time periods. The data are from measurements at Stage I of Hurricane Betsy. The number of datasets plotted is a matter of choice. A quadratic polynomial curve was fit to the data by the method of least squares. The curve obtained is shown in the figure. The fitted curve is defined by the equation

$$\gamma = 3.724 - 17.26k + 68.32k^2 \tag{15}$$

Once $\gamma$ and the slope $d\gamma/dk$ are determined for a given wave period equation (14) is used to determine $\phi$ and then equation (11) to determine the wave packet bearing at the measurement site. Note that it is not necessary to know the water depth where the waves were generated using this method.

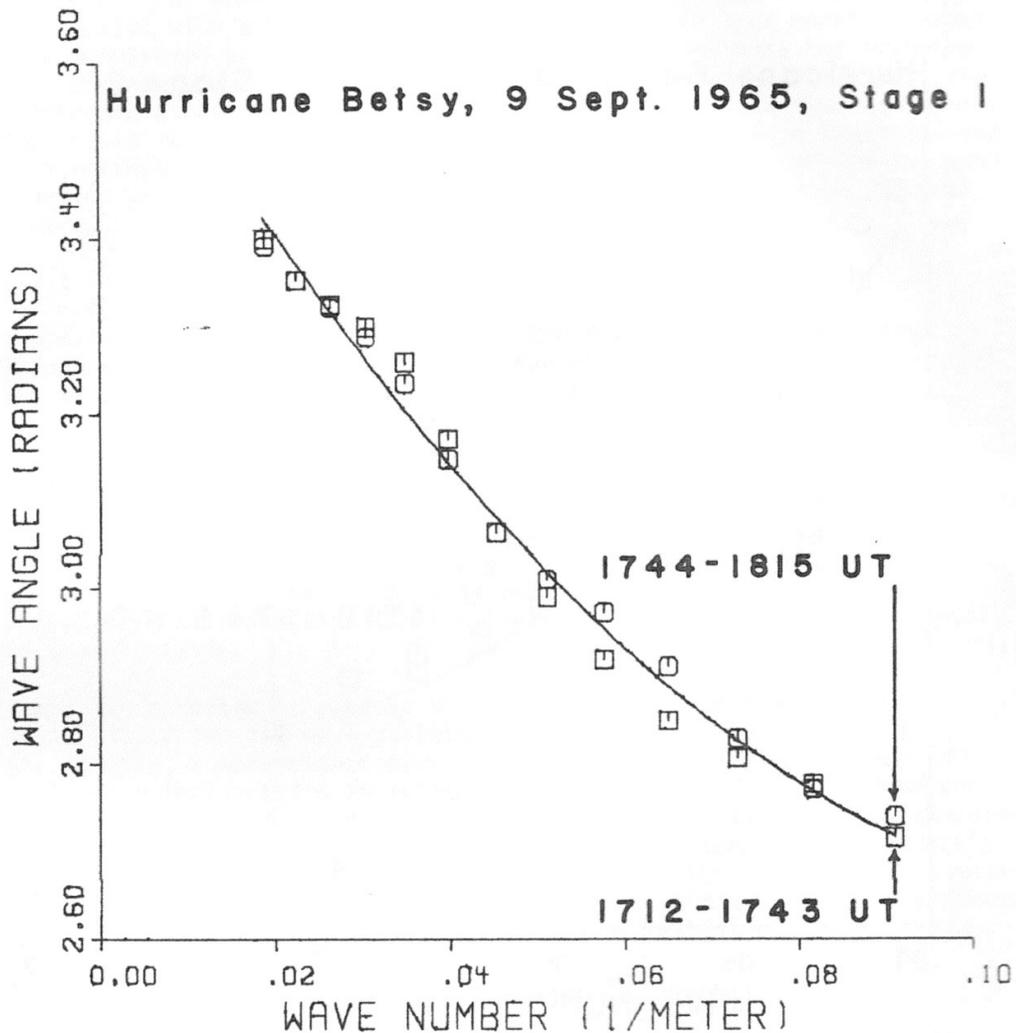

**Figure 4:** Wavelet directions as a function of wave number. The data are from Hurricane Betsy. The times are Universal Time.

## 9.2. Hurricane Fifi

A refraction example is presented by Breeding (1978) for waves measured near Panama City, Florida with the array of pressure sensors at Stage I described in sections 4.1 and 4.2. The swell waves were from Hurricane Fifi which passed through the Caribbean Sea off the coast of Honduras in September 1974.

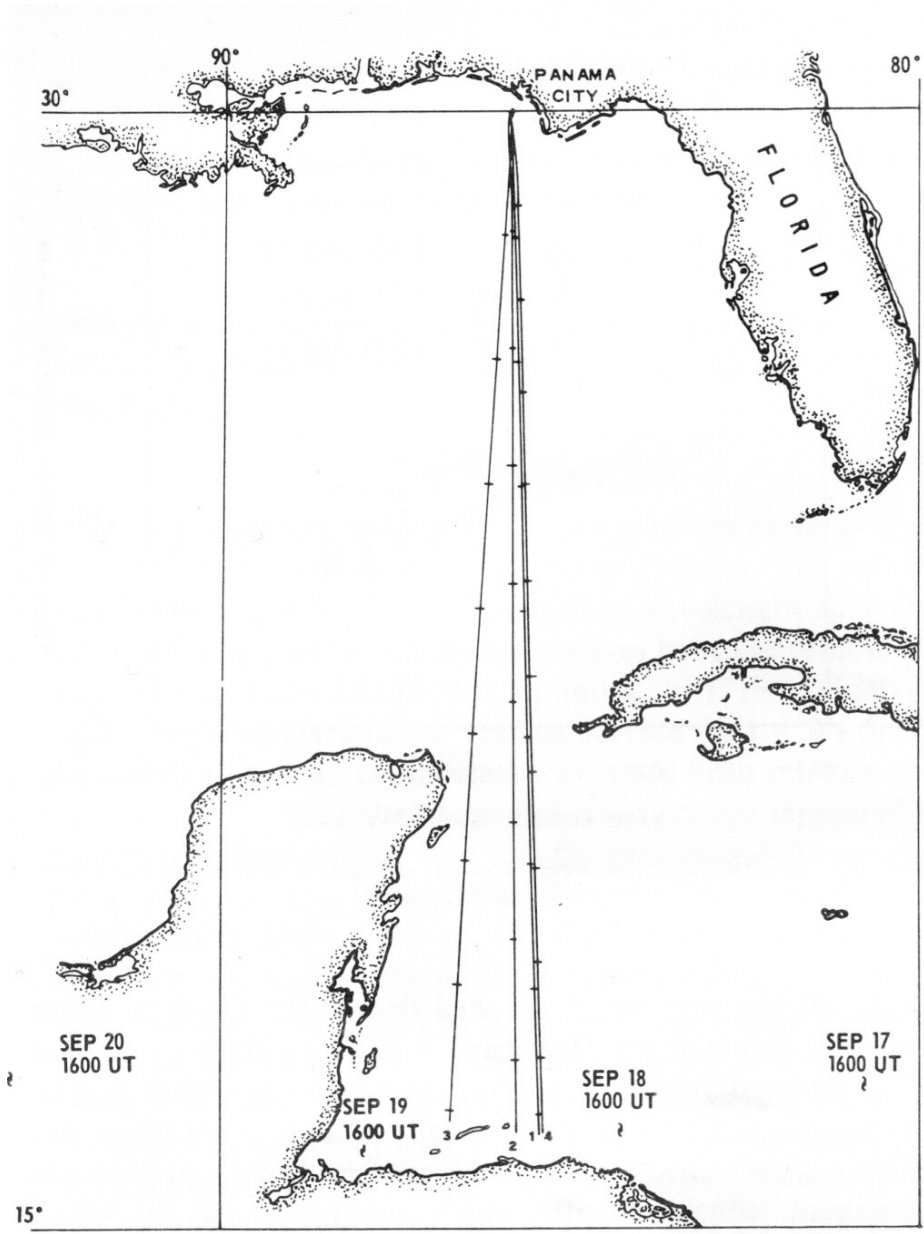

**Figure 5:** Backtracks for Hurricane Fifi from Stage I. Rays 1 and 2 were from data measured on September 19, 1974 from 2250-2315 Universal Time. Rays 3 and 4 were from data measured on September 20. 1974 from 0850-0915 Universal Time. Rays 1, 2, 3, and 4 have wavelet periods of 16, 15, 16, and 12 sec and measured wavelet bearings of 195.82, 194.05, 203.34, and 197.35 degrees, respectively. The 12 sec wave period hydrons moves quite a bit slower than the other wave period hydrons taking longer to reach the measurement site. The path of Hurricane Fifi is shown on the bottom of the figure.

To determine the wave packet directions for the different wave periods both the refraction-refraction source method and the direction-wave number method were used. The backtracks for the wave packet are shown in figure 5. It can be seen that the wave packet rays go back through the Yucatan Channel to the storm. Backtracks for monochromatic waves with phase velocity (Breeding, 1978) miss the Yucatan Channel and instead strike land on both sides of the channel.

This is another example of using two refraction laws to refract hydrons. The hydrons refract according to the ray curvature expression with the geometric group velocity while the wavelets within a hydron refract according to Snell's law with phase velocity. For more details including the illustration of the monochromatic backtracks see Breeding (1978).

## 9.3. Backtracks to Distant Storms - Continued

Munk, et al (1963) obtained poor results with their inferred backtracks to distant storms based on Snell's law with phase velocity described in section 3.3. As a result, it is important to determine backtracks for their data using the two refraction laws for propagating hydrons. My master's student Tang (1994) determined both monochromatic and hydron backtracks for the bathymetry in the vicinity of the measurement site. He created a water depth grid by taking the actual water depths from bathymetric charts. The spacing between grid points is 1.85 km.

Four storms were selected for study. One storm had an origin of 17.1 August 1959 where 17.1 is the day and time. Here .1 is the hour, e.g., .5 is noon. This storm was located nearly half-way around the world from the measurement site. It was antipodal swell from the Indian Ocean. For two of the storms, 4.5 and 11.6

September 1959, their wave-inferred sources were on the Antarctic Continent. The remaining storm of 8.4 October 1959 was located in the Southern Ocean.

For each of the selected storms backtracks were determined by Tang (1994) for a number of datasets recorded at different times. To produce the backtracks a total of 13 datasets were chosen. The wave periods in this study varied from 17 to 27 seconds.

The hydron backtracks were determined following Breeding (1978). From the measured wavelet bearings the hydron bearings were determined by the refraction-source method defined in section 6.2.1. Hydron backtracks were computed using the ray curvature expression with the geometric group velocity. At each ray point the wavelet directions were evaluated using Snell's law with phase velocity.

Do to the large number of refraction diagrams the results are summarized in Table 1. The Weather Map gives the bearings to the storms considered. The wave bearings in degrees presented are the deep-water values. The findings of Munk, et al (1963) are included for comparison. Although the larger wave periods which are affected more by refraction were ignored in the Munk et al (1963) study, they were included in this investigation. The Monochromatic and hydron backtracks refer to the spread in bearings of the computed backtracks found by Tang (1994). The angle $\Delta$ is the distance from the measurement site to the storm. A value of 180 degrees would be halfway around the world.

**Table 1:** Hydron and Monochromatic Backtracks for Southern Hemisphere Storms.

| Storm | 17.1 Aug | 4.5 Sep | 11.6 Sep | 8.4 Oct |
|---|---|---|---|---|
| Weather Map | 210 - 220 | 214 | 215 | 220 |
| Munk, et al (1963) | 223 | 200 | 200 | 215 |

| Monochromatic | 216 - 260 | 206 - 242 | 210 - 242 | 207 - 240 |
| --- | --- | --- | --- | --- |
| Hydrons | 220 - 223 | 220 - 223 | 220 - 223 | 220 - 223 |
| Distance Away | Δ = 175 | Δ = 135 | Δ = 126 | Δ = 139 |

The monochromatic backtracks for all of the storms badly diverge and do not go back to a common source.  Many of the monochromatic backtracks miss the narrow windows that they must go through to reach the storm generation areas.

All of the hydron backtracks for each storm are consistent with a common source and were found to go back within the angular range of 220 to 223 degrees.  This range is consistent with the 216 to 225 degrees window south of New Zealand when limited by Antarctic pack ice.  The two storms which Munk, et al (1963) located on land in their monochromatic ray analysis were found to be located at sea when considering hydron backtracks.  Further, whereas the Munk, et al (1963) analysis led to inferred bearings to the left of the storm centers when viewed from the measurement site, the hydron inferred storm locations are to the right of the storm centers located on weather maps.  This is consistent with the clockwise wind pattern found in the southern hemisphere.  These results by Tang (1994) clearly show that wave packets refract according to Snell's law with the geometric group velocity while the wavelets within the packet refract according to Snell's law with phase velocity.  The wave packets do not refract the same way as monochromatic waves.

In view of the good results presented here the Munk et al (2013) article in which they propose a correction to their 1963 article, discussed in section 3.3, needs to be reevaluated.  The correction is based on the article by Gallet s Young (2014) where they propose that rays for some of the storms are deflected away from great-circle paths by refraction due to the vorticity of currents. But the corrections, to the extent they are needed, should be made to the hydron backtracks discussed here.

## 10. Attributes of Propagating Wave Packets

### 10.1. Properties of the Hydron Ray Curvature Expression

Breeding (1980) derived a ray curvature expression for hydrons that can be stated

$$\kappa_G = \frac{\frac{1}{U}\frac{dU}{dx} + \frac{tan\phi\ tan\gamma}{v}\frac{dv}{dx}}{csc\theta + tan\phi\ sec\theta} \quad (16)$$

This equation is evaluated in the primed coordinate system where the first order derivatives of y vanish. This equation can be used in a hydron prediction program.

Equation (16) has some remarkable and very interesting properties for propagating hydrons. To simply the discussion the water depth contours are taken parallel to the y-axis. It is assumed that the velocities and their derivatives are continuous and finite. However, under some conditions the trigonometry terms become infinite or have indeterminate forms. For example, the ray curvature approaches zero if the hydron direction $\theta$ becomes either parallel or perpendicular to the wave speed contours, provided the wavelet direction $\gamma$ is not parallel to the same contours. This means that for a sufficiently long path refraction turns the direction of a hydron so that it is either parallel or perpendicular to the wave speed contours. The first case is seen in Figure 3 for the hydron trajectories which are nearly parallel to the water depth contours along the west coast of Florida.  This would not be possible for monochromatic rays that refract according to Snell's law with phase velocity.

## 10.2. Critical Angle for Hydrons Propagating from Deep Water

Breeding (1980) considered hydron trajectories where the water depth contours are parallel and the hydrons propagate from deep water towards shore.  Initially the hydron and wavelet angles are equal. The angles are measured with respect to the normal to the water depth contours.  Regardless of the wave period, for deep water incident angles between 0 and 74.8 degrees the hydrons follow paths such that the angles increase to the depth of the geometric group speed maximum, then undergo a point of inflection, and then decrease shoreward.  As the hydron approaches shore its direction becomes perpendicular to the wave speed contours and the hydron ray curvature approaches zero. For deep water angles of incidence equal to or greater than 74.8 degrees refraction will turn the hydron direction parallel to the shoreline at an intermediate water depth and the hydron will continue moving in that direction.  This is another example of the ray curvature going to zero when the hydron direction becomes parallel to the wave speed contours.  The angle 74.8 degrees is a critical angle.  Rays propagating from deep water are seen in Figure 6.

Breeding (1980) also considered the maximum percentage difference of G from U for the incident angles in Figure 6.  When the initial deep-water angle $\theta$ = 30 degrees the maximum percentage difference is 2.70%, when $\theta$ = 45 degrees the value is 5.91%, for $\theta$ = 60 degrees the value is 10.03%, and for $\theta$ = 74.8 degrees the value is 14.27%.  Point by point along a ray the percentage difference of G from U is usually much less then these maximum values.  This explains why the refraction of hydrons with Snell's law with the conventional group speed U instead of G, or the ray curvature expression in U instead of G, can yield a close approximation to the correct answer.

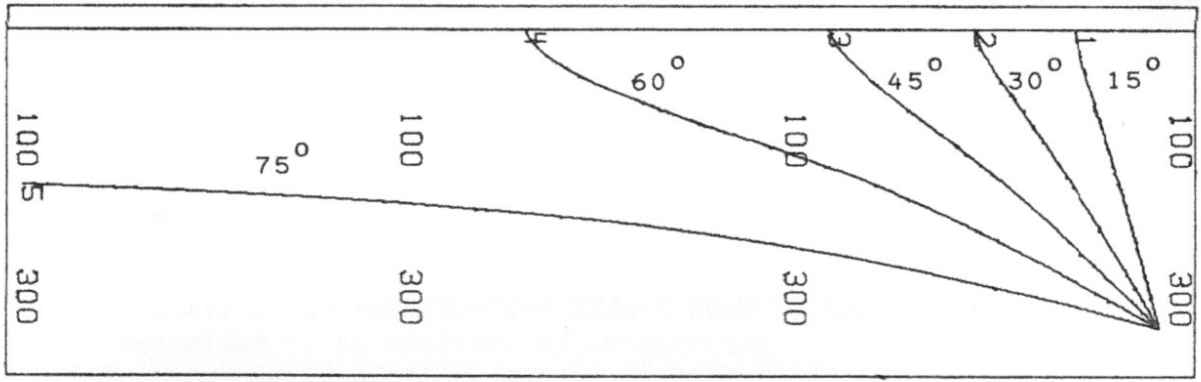

**Figure 6:** Critical angle for hydron trajectories propagating from deep water. The illustration is for a 20-second wave period. The water depth contours are parallel and the water depths are in meters. The initial hydron direction is shown for each ray and is the angle between the hydron velocity vector and the normal to the water depth contours. The critical angle is 74.8 degrees. For that angle and greater angles, regardless of wave period, the trajectory is turned parallel to the shoreline at an intermediate water depth.

## 10.3. Total Reflection for Hydrons Propagating towards Deep Water

Breeding (1980) shows that If the hydron direction is neither parallel nor perpendicular to the wave speed contours, then if the wavelet direction becomes parallel to the wave speed contours the ray curvature becomes infinite. This can be seen in equation (16). It can happen if the hydrons start in shallow water or an intermediate water depth and propagate toward deeper water. If refraction turns the wavelet directions parallel to the water depth contours the hydrons undergo total reflection. As the reflection point is approached and the hydron ray curvature approaches infinity the direction of a hydron becomes perpendicular the water depth contours. Further, at the reflection point the hydron velocity goes to zero, which is consistent with a particle concept.

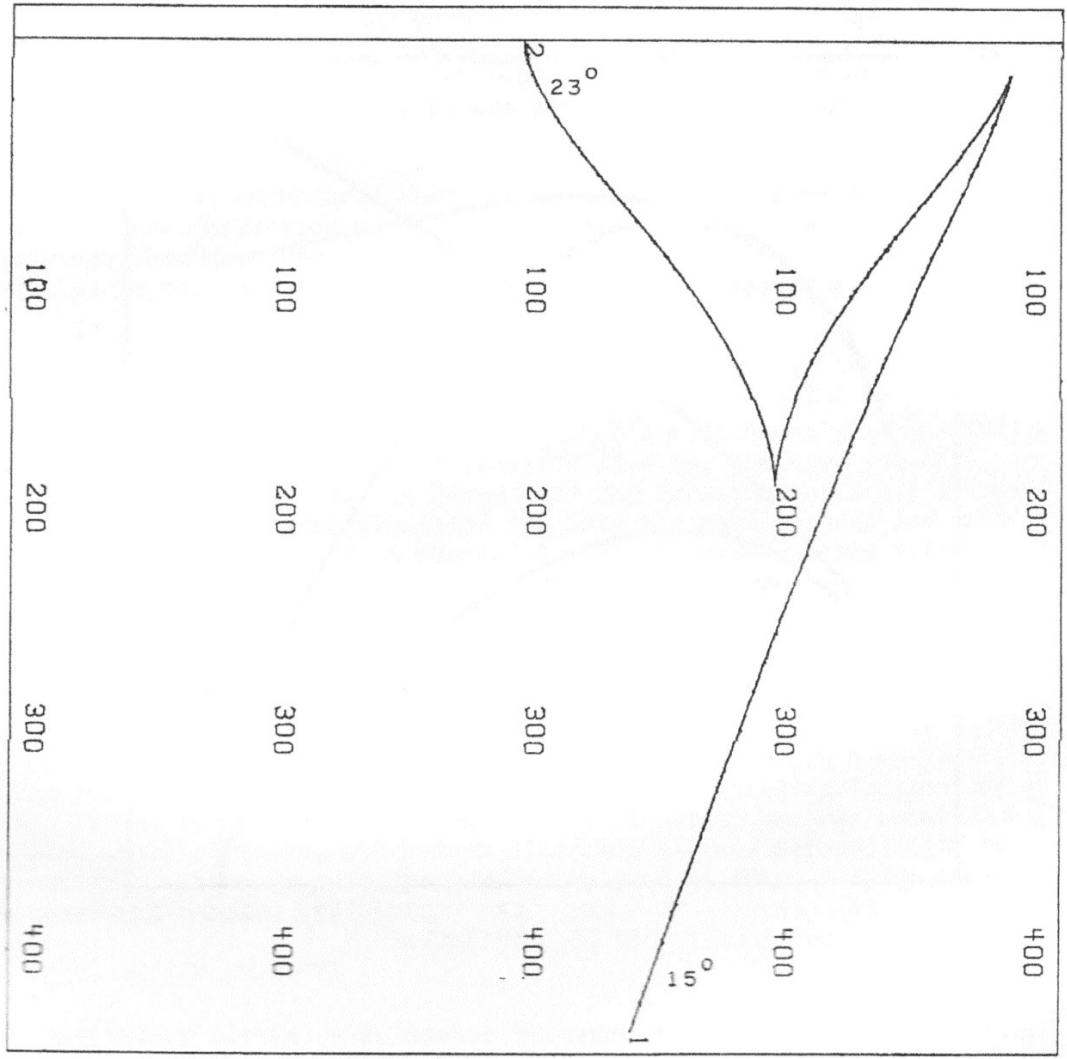

**Figure 7:** Reflection point. The hydron trajectories are for 20 second wave periods and the waves begin at an intermediate water depth. The water depths are in meters. The initial hydron direction is shown for each ray and is defined as in Figure 6.

As in quantum mechanics the wave-particle duality is encountered. A reflection point is seen for one of the rays in Figure 7. The other ray does not reflect due to its angle of incidence.

## 10.4. Fermat's Principle, Euler-Lagrange Equation, and Ray Curvature

Fermat's principle played a vital part in my decision that backtracks should be determined by Snell's law or a ray curvature expression with a group velocity. It is beyond the scope of this article, but Breeding (1986) shows that Fermat's principle can be used to derive the Euler-Lagrange equation. This equation determines the path of minimum travel time between two points. and the ray curvature expression is derived from it. By integrating the ray curvature expression with the geometric group velocity Snell's law with the geometric group velocity is derived. This provides a theoretical basis for wave packet refraction.

## 10.5. Hamilton's Equation and Classical Mechanics

Breeding (1986) shows that by starting with Hamilton's equation in classical mechanics an expression can be derived that looks like the ray curvature expression. There is a strong analogy between particles in classical mechanics and the rays of waves. The wave-particle duality in quantum mechanics holds that every particle can also be described in terms of waves. The reverse has been shown here. The wave packets of waves also have a particle aspect.

## 10.6. Hydrons and Alongshore Migrating Undulating Beach Forms

For sinuous water depth contours with a sinuous beach shoreline refraction according to Snell's law with phase velocity has more energy at the headlands and less energy in the bays. It would seem that there should be more energy in the bays where land is missing. Breeding (1981) considered the refraction of hydrons from deep water for sinuous water depth contours. More wave energy was found in the bays near shore where the rays converge than at the headlands where the rays diverge. For a set of beach undulations Sonu (1972) observed higher water wave heights in the bays than at the headlands, which supports the refraction of hydrons with the two refractions laws rather than the refraction of monochromatic waves.

Breeding (1981) also shows that a beach undulation can be an equilibrium beach form where the combined refraction of the hydrons and the wavelets within the hydrons provide a constant longshore drift of sediments.

## 11. Wave Tank Test of Hydron Refraction

It was possible to test the refraction of gravity water wave hydrons using a wave tank available to the Florida Institute of Technology in Melbourne. The wave tank is at the Indian River Marine Science Center located at Vero Beach south of the Melbourne campus.

The obvious thing to test is the critical angle of 74.8 degrees for hydrons propagating from deep water. At incident angles equal to or greater than the critical angle the trajectories would refract and turn parallel with the water depth contours. They would not reach shallow water. This is seen in Figure 6. If instead the hydrons refract according to Snell's law with phase velocity as monochromatic waves the waves would travel into shallow water. The objective is to see which of these theories is correct by examining where the waves go in the wave tank. My master's student Linzell (1987) performed the test.

The wave tank is a rectangular structure of cinder blocks placed on a concrete slab floor. The dimensions are 4.88 m by 3.96 m with a maximum water depth of 0.762 m. The wave tank has a number of different kinds of wave absorbers to reduce reflections and standing waves on all sides of the tank.

There are two different water bottoms used to exploit the refraction of the dispersive water waves in intermediate water depths. For some testing of the refraction theories the water

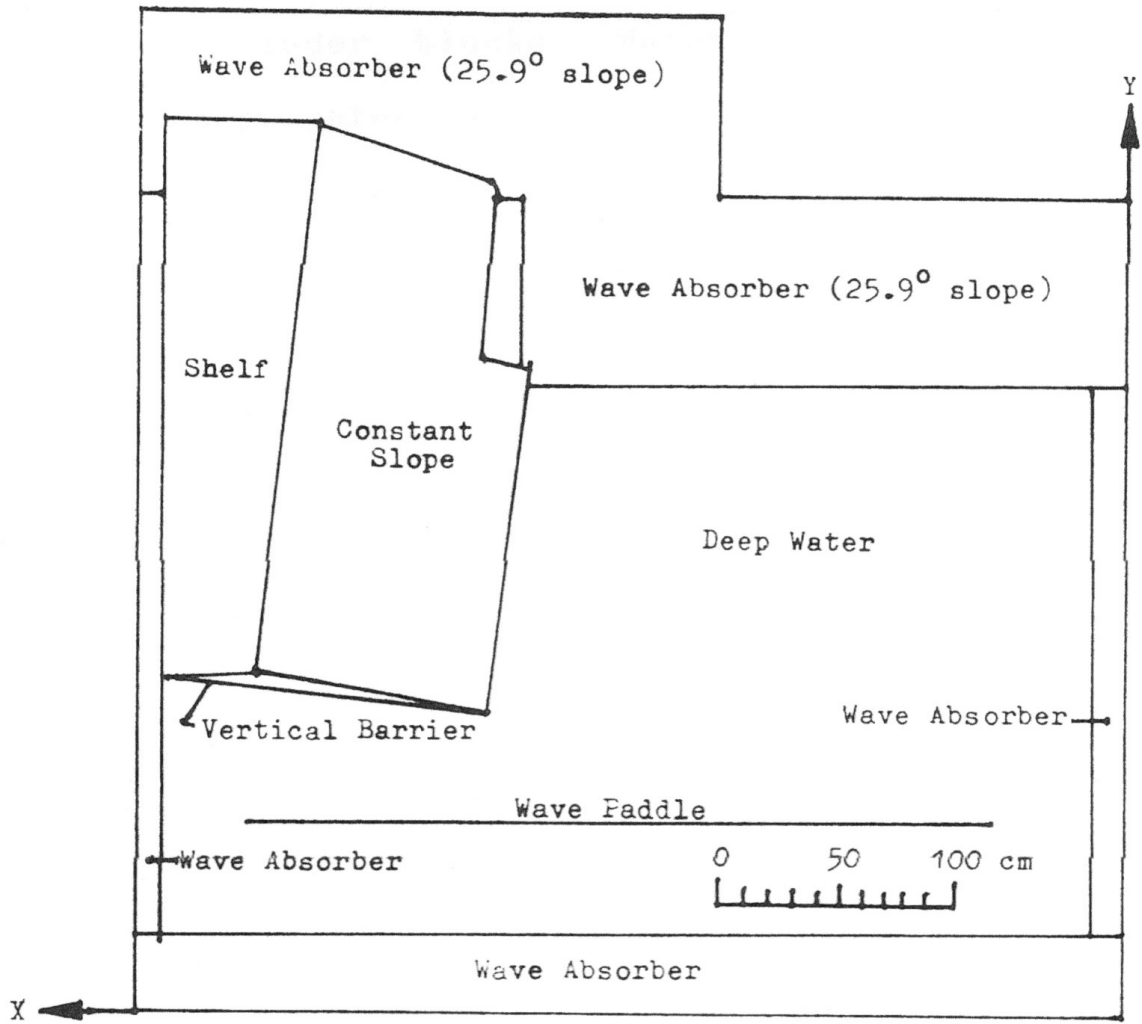

**Figure 8:** Wave tank plan view. The depth in deep water is 33.4 cm. The depth in shallow water (shelf) is 0.40 cm with the step discontinuity and 0.56 cm without it. The inclination of the slope is 6.15 degrees with the step discontinuity and 16.71 degrees without it. Note that in order to generate waves at the critical angle the wave paddle is at an angle of 74.8 degrees with respect to the slope water depth contours.

bottom included a step discontinuity. The plan of the wave tank is seen in Figure 8.

Waves are generated at the deep-water end of the tank.  The wave maker is a flap-type paddle hinged at the bottom and driven by an electric motor.  There are two motor speeds.  Wheels are attached to the left bottom of the wave maker so that waves can be generated at different angles of incidence in deep water.  Four incident angles were used to test the refraction laws: 72, 74, 76, and 78 degrees.  To generate wave groups the wave maker was turned on for five complete cycles before it was turned off.

The following wave periods can be generated in the tank.  For a motor speed of 1140 RPM: 0.65, 0.82. 0.95, 1.17, and 1.36 seconds.  For a motor speed of 1720 RPM: 0.43, 0.54, 0.60, 0.79, and 0.98 seconds.  For the wave period chosen the water depth must be greater than one half the wavelength so that the waves generated are in deep water.  The wavelengths must also be small compared to the dimensions of the tank.

A wave staff was used by Linzell (1987) to measure the waves and it was placed at different locations in the tank to determine where waves existed and how big they were.  The data from the wave staff was recorded on a strip chart recorder.

In addition, pictures of the waves generated were taken from a position above the tank with a 35 mm camera at a shutter speed of 1/1000 of a second to freeze the wave pattern.  The wave patterns generated were also recorded using a video tape recorder.  The three methods of measurement produced similar results.

Figure 9 (Linzell, 1987) shows an example of the results obtained from the wave measurements.  The wave staff records are shown for an incident angle of 74 degrees and a wave period of 0.598 seconds.  There are 6 records.  The record at the bottom of the figure is for the wave staff in deep water.  For each successive record going towards the top of the figure the intermediate

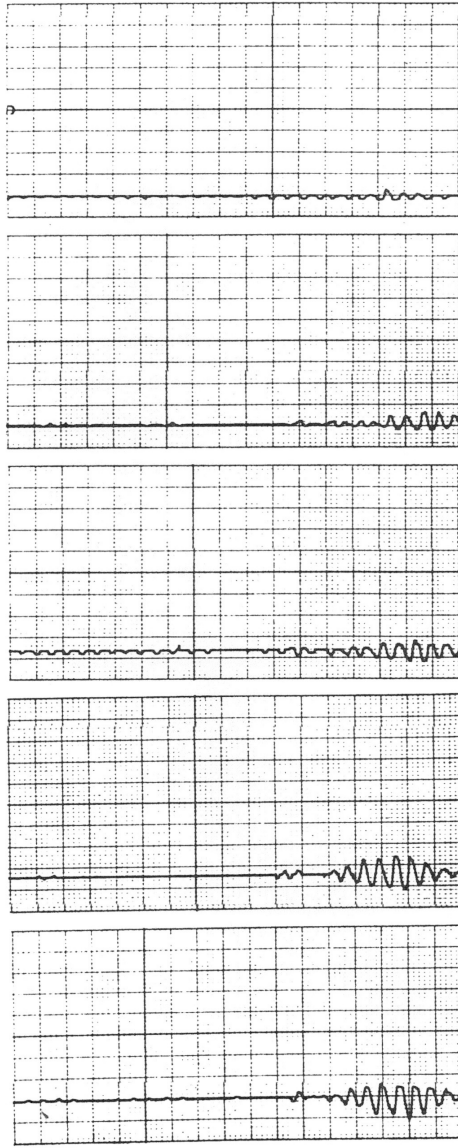

**Figure 9:** Wave staff records across the sloping bottom for model runs. The wave staff record at the bottom of the figure is in deep water. For each successive wave staff record going towards the top of the figure the intermediate water depths are decreasing. The wave heights of the measured waves are seen to decrease as the water depth decreases.

water depth is decreasing. The wave staff records clearly show a large decrease in wave energy going across the sloping bottom.

Very little if any wave energy reaches the shelf in shallow water. This is clearly inconsistent with the refraction of monochromatic waves according to Snell's law with phase velocity. Instead, the results offers strong support for the refraction of hydrons according to Snell's law with the geometric group velocity, while at each ray point the wavelet direction is determined by Snell's law with phase velocity.

## 12. Conclusions

Wave patterns considered in time for a constant position versus the patterns in distance for an instant of time are different if the waves are dispersive. The difference is related to the ratio of the conventional group speed U to the phase speed v.

In dispersive media the speed of a wave packet is given by the geometric group speed $G = U cos(\theta - \gamma)$ where $U = d\omega/dk$, $\theta$ is the direction of the wave packet, and $\gamma$ is the wavelet direction.

The refraction of a wave packet requires two refraction laws. The direction of the wave packet is determined by Snell's law with the geometric group velocity, or the equivalent ray curvature expression. At the same time the direction of the wavelets within the wave packet is determined by Snell's law with the phase velocity.

High quality directional wave data were used to test the refraction laws. In all cases the directional wave data were measured with arrays of pressure sensors. The direction measured is the wavelet bearing to the storm generational area. The hydron bearing has to be determined. This can be done using the refraction-source method or the direction-wave number method, both of which are described.

A total of 15 datasets were tested and are described. This includes data from storms in the Caribbean Sea, Gulf of Mexico, Southern Ocean, and the Indian Ocean. In all cases, hydron (wave packet) refraction, as defined above in terms of two refraction laws produced very good results with backtracks for different wave periods going back together to each storm. This includes the larger wave periods which undergo large amounts of refraction. For the same datasets the monochromatic backtracks spread out and are not consistent with common sources

The use of the conventional group velocity **U** in place of the geometric group velocity **G** leads to simplifications in the calculations and often yields a good approximation to the correct answer. In computing some of the backtracks **U** was used in place of **G**.

For hydrons moving from deep water there is a critical angle of 74.8 degrees. For angles less than that if the water depth contours are parallel all of the hydrons reach shore where the hydron direction becomes perpendicular to the water depth contours. But for a deep-water angle of incidence equal to or greater than 74.8 degrees refraction will turn the hydron direction parallel to the shoreline at an intermediate water depth, and the hydrons will continue moving in that direction. This would not be possible for a monochromatic wave.

The refraction laws were tested in a wave tank with a test of the critical angle of 74.8 degrees for wave groups generated in deep water. The waves did not reach shallow water as they would if they refracted according to Snell's law with phase velocity. The results strongly verify that wave groups refract according to Snell's law with the geometric group velocity and not as monochromatic waves with phase velocity.

Total reflection can occur for a hydron if it starts in shallow water or an intermediate water depth and propagates toward deeper water. If refraction turns the wavelet directions parallel to the water depth contours the hydron undergoes total reflection. At the reflection point the hydron direction becomes perpendicular to the water depth contours and the hydron velocity goes to zero similar to a particle. The wave-particle duality is encountered

For undulating beach forms the refraction of hydrons can yield more energy in the bays and less energy at the headlands. This is just the opposite of what happens for monochromatic wave refraction. Field data has been reported which supports hydron refraction with larger wave heights in the bays.

## 13. Acknowledgments

I thank Jacob George for reviewing the manuscript. I thank Kenneth C. Matson and Nourollah Riahi for their help in creating a computer program to compute the trajectories and wave heights of hydrons.